\documentclass[twocolumn,floats,showpacs,prd,amsmath,amssymb,floatfix,nofootinbib,balancelastpage]{revtex4}
\input epsf
\usepackage{graphics}
\usepackage{amsmath}
\usepackage{color}
\usepackage{dcolumn}

\usepackage{graphicx}

\newcommand{\beqa}{\begin{eqnarray}}
\newcommand{\eeqa}{\end{eqnarray}}

\newcommand{\bn}{\hat{\bf n}}
\newcommand{\bl}{\hat{\bf l}}
\newcommand{\beq}{\begin{equation}}
\newcommand{\eeq}{\end{equation}}
\newcommand{\bfl}{{\mathbf{l}}}

\newcommand{\bfL}{{\mathbf{L}}}
\newcommand{\bfLp}{{\mathbf{L^{\prime}}}}
\newcommand{\bflp}{{\mathbf{l^{\prime}}}}

\newcommand{\intl}[1]{\int {d^2 l_{#1} \over (2\pi)^2}}
\newcommand{\intlp}[1]{\int {d^2 l_{#1}' \over (2\pi)^2}}
\newcommand{\vsp}{\vphantom{\Big[}\\}

\newcommand{\obs}{{\rm obs}}

\newcommand{\vecl}{{\bf l}}
\newcommand{\vecla}{{{\bf l}_1}}
\newcommand{\veclb}{{{\bf l}_2}}

\newcommand{\dirac}{{\rm D}}

\newlength{\tskip}
\newlength{\colwidth}

\newcommand{\len}{\phi^{len}}

\newcommand{\estEB}{{{\hat \alpha}}_{EB}}

\newcommand{\normEB}{A_{EB}}

\newcommand{\filtEB}{F_{EB}}

\newcommand{\noiseEB}{N_{EB,EB}}

\newcommand{\noiseLEB}{N^{(len)}_{EB,EB}}
\newcommand{\noiseAEB}{N^{\alpha\alpha}_{EB,EB}}
\newcommand{\noiseXEB}{N^{(X)}_{EB,EB}}

\newcommand{\noiseWEB}{N^{\omega\omega}_{EB,EB}}

\newcommand{\vl}{{\mathbf{l}}}

\begin{document}
\title{
Constraining a spatially dependent rotation of the Cosmic Microwave Background Polarization}

\author{Amit~P.S.~Yadav$^1$,
        Rahul~Biswas$^2$,
        Meng~Su$^1$, and
        Matias~Zaldarriaga$^{1,3}$}

\affiliation{$^1$ Center for Astrophysics, Harvard University, Cambridge, MA 02138, USA}
\affiliation{$^2$ Department of Physics, University of Illinois at Urbana-Champaign, Urbana, IL 61801, USA}
\affiliation{$^3$ Department of Physics, Harvard University, Cambridge, MA 02138, USA}

\begin{abstract}
Following Kamionkowski (2008), a quadratic estimator of the rotation of the plane of polarization of the CMB is constructed. This statistic  can estimate a spatially varying rotation angle $\alpha(n)$. We use this estimator to quantify the prospects of detecting such a rotation field with forthcoming
experiments. For  PLANCK and CMBPol we find that the estimator containing the product of the $E$ and $B$ components of the polarization field  is the most sensitive. The variance of this EB estimator, $N(L)$ is roughly independent of the multipole $L$, and is only weakly dependent on the instrumental beam. For FWHM of the beam size $\Theta_{fwhm}\sim 5'-50'$, and instrument noise $\Delta_p \sim 5-50 \mu K$-arcmin, the scaling of variance $N(L)$ can be fitted by a power law $N(L)=3.3\times 10^{-7} \Delta^2_p \Theta^{1.3}_{fwhm}$ deg$^2$. For small instrumental noise $\Delta_p \leq 5 \mu K$-arcmin, the lensing B-modes become important, saturating the variance to $\sim10^{-6}$deg$^2$ even for an ideal experiment. 
Upcoming experiments like PLANCK will be able to detect a power spectrum of the 
rotation angle, $C^{\alpha \alpha}(L)$, as small as $0.01$ deg$^2$, while futuristic experiment like CMBPol will be able to detect rotation angle power spectrum as small as $2.5 \times 10^{-5}$ deg$^2$.  We discuss the implications of such constraints, both for the various physical effects that can rotate the polarization as photons travel from the last scattering surface as well as for constraints on instrumental systematics that can also lead to a spurious rotation signal. Rotation of the CMB polarization generates B-modes which will act as contamination for the primordial B-modes detection. We discuss an application of our estimator to de-rotate the CMB to increase the sensitivity for the primordial B-modes.

\end{abstract}
\maketitle
\section{Introduction}
The polarization of the Cosmic Microwave Background (CMB) field can be studied in terms of the parity even E and parity odd 
B-modes~\cite{SZ97,SZ98,1997PhRvD..55.7368K,1997PhRvL..78.2058K}. 
In standard cosmology, the physics governing the radiating field is parity 
invariant. Hence, the parity odd correlations
$\langle T B \rangle$, $\langle E B \rangle$ vanish identically irrespective of the exact values of the cosmological parameters. However, the plane of linear polarization of CMB fields can be 
rotated due to interactions which introduce a different 
dispersion relation for the left and right circularly 
polarized modes, during propagation from the surface of last 
scattering surface to the earth. 
Such rotations generate non-zero cross-correlations $\langle T B \rangle$, $\langle E B \rangle$ in the CMB field.
Thus, measurement of these correlations allows us to estimate the 
rotation of the plane of the CMB polarization~\citep{1999PhRvL..83.1506L}.
Such interactions can come from three
main sources: (a) interaction with dust foregrounds, (b) Faraday rotation 
due to interaction with background magnetic fields, and (c) interactions with 
pseudoscalar fields~\cite{Carroll98}. The interaction with foregrounds leads to a  
a frequency dependent effect. The same is true of Faraday rotation, where the frequency dependence
is ($\sim \nu^{-2}$) ~\citep{Kosowsky_Loeb1996,2005PhRvD..71d3006K, 2004ApJ...616....1C,2004PhRvD..70f3003S}, while the interaction with pseudo-scalar fields is frequency
independent. The distinct frequency dependencies allow one to separate these effects.


We know that parity is violated by weak interactions, and is possibly violated in the early universe, to give rise to baryon asymmetry. 
Hence, investigating the existence of parity violating interactions involving 
cosmologically evolving scalar fields is well motivated.  As an example we consider an interaction of the form $\frac{\phi}{2M}F_{\mu \nu}{\tilde F}^{\mu \nu}$~\cite{Carroll98,prs08}. 
It has been shown that such a term can rotate polarization vector of linearly polarized light by an angle of rotation 
$\alpha = \frac{1}{M}\int d\tau \dot{\phi}$ 
during propagation for a conformal time $\tau$. The fluctuations in the scalar field $\phi$ then will be imprinted in the rotation angle $\alpha$ of the polarization.
It is interesting to ask what is the level of these fluctuations that can be detected with the upcoming CMB polarization experiments. 

However, the observational 
situation is somewhat complicated by the fact that the 
measured CMB fields could be rotated with respect to the signal due to 
instrumental systematics:
a mis-calibration of the orientation of the instrument which results in a 
constant rotation and differential offsets of the orientation of the 
individual detectors of the instrument resulting in rotation dependent on
angular position $\bn$.
It should be noted that this systematic effect is also a concern for the 
detection of polarization B modes, a major goal for forthcoming 
polarization experiments, since it can result in a spurious B-mode
 detection.
Rotation of the CMB $\alpha({\bf n})$, either due to interactions with a 
pseudoscalar field, or due to the instrumental rotation miscalibration can be a function of angular position ($\bn$).

An estimator for the spatially varying ration angle  
$\alpha(\bn)$ has been reported in ~\citet{2008arXiv0810.1286K}. 
Here, we point out that the widely used formalism for gravitational 
lensing~\citep{2002ApJ...574..566H} may be suitably modified to describe the 
rotation of CMB polarization. We use this to construct approximate, but 
simpler form of the quadratic estimator of 
$\alpha(\bn)$ using the flat-sky limit, and study its variance. 

These 
estimators may be used to study the physics behind the rotation of the
polarization of light. 
As we shall discuss, we can also put an upper-bound on the 
frequency independent rotation from non-standard interactions. We also discuss the use our the estimator $\hat{\alpha}(\bn)$ to control instrumental rotation 
systematics for the detection of primordial B-modes. The presence of rotation systematics in the instrument
generates B-modes. Hence, the control of rotation systematics
of instruments is 
important for the measurement of B-modes of the CMB polarization; a major 
goal of subsequent polarization experiments. In this paper, we will 
discuss the prospects of detecting non-standard physics through measuring
the angle of rotation, and the level to which we can control rotation 
systematics by this estimator. 

A search for a constant rotation of the polarized light by an angle 
$\alpha$ from radio galaxies and the CMB is already underway~\cite{wmap5_cosmology,quad_parity,2008ApJ...679L..61X,2008A&A...483..715X,2006PhRvL..96v1302F,2008PhRvD..78l3009K,Carroll89,2007PhRvD..76l3014C}. So far, 
there is no evidence of non-zero angle of rotation, and the angle 
$\alpha$ is constrained to be less than a few degrees 
~\cite{wmap5_cosmology,quad_parity}. At present, there are no studies of constraints on spatially varying rotation angle $\alpha (\hat{n})$.

\section{Formalism}
In this section, we will construct an estimator for the spatially varying rotation field. We shall also describe a physical scenario which give rise to a 
frequency independent but spatially varying rotation.
\subsection{Estimator for Spatially Dependent Rotation Field}
\label{sec:formalism_estimator}                                
We will describe the observable effect of rotation on 
the CMB polarization fields. Let the un-rotated (usual) CMB temperature 
field and the Stokes parameters  at angular position $\bn$  be
$\tilde{T}(\bn)$, and $\tilde{Q}(\bn)$, $\tilde{U}(\bn)$ respectively.
The relevant ensemble averages of the un-rotated CMB field can be
encapsulated in
\begin{equation}
\left\langle \tilde{x}(\bfl)\right\rangle = 0, \qquad
\left\langle \tilde{x}^\star(\bfl)\tilde{x}^{\prime}(\bfl^{\prime})\right\rangle
=
(2\pi)^2 \delta(\bfl -\bfl^{\prime})\tilde{C}_{\bfl}^{xx^{\prime}}\,,
\end{equation}
where $x,x^{\prime}$ run over the $T, E$, or $B$ fields, and $\tilde{C}_{\bfl }^{xx^{\prime}}$ is the un-rotated CMB power spectrum.
The temperature fields are invariant 
under a 
rotation of the polarization by an angle $\alpha(\bn)$ 
at the angular position $\bn$,
while the Stokes parameters transform like a spin two field. Thus, due to 
rotation, the observed fields are
\begin{equation}
(Q(\bn) \pm iU(\bn))
=(\tilde{Q}(\bn) \pm \tilde{U}(\bn)) \exp(\pm 2i\alpha(\bn))
\label{RotationTransformationOfStokes}
.
\end{equation}
 The $E$ and $B$ fields of the CMB can be constructed from 
 observed Stokes parameters. In a Fourier basis, in the flat sky approximation,
\begin{eqnarray}
 \left[ E\pm i B \right] (\bfl) &=&
        \int  d \bn\, [Q(\bn)\pm i U(\bn)] e^{\mp 2i\varphi_{\bf l}} e^{-i \bl \cdot \bn}\,,
\label{EBFields}
\end{eqnarray}
where $\varphi_{\bfl}=\cos^{-1}(\hat {\bf n} \cdot \hat \bfl)$. Since, the angle of rotation is already constrained to be small, we will 
work out the effects to first order in $\alpha(\bn)$. 

Since, we can only compute correlations of the CMB polarization modes 
theoretically, we want to isolate the change in correlations due to this 
rotation. 
Even in the absence of the physics causing rotation, we expect the CMB 
fields to be gravitationally lensed by matter inhomogeneities. 
Hence, the change due to rotation is the difference between lensed 
rotated fields, and lensed un-rotated fields $\tilde{T},\tilde{E},\tilde{B}$.
We make the similarity of our problem with gravitational lensing
of the CMB~\cite{2002ApJ...574..566H}  manifest by writing the
change in the CMB field modes $\delta x(l) = x(l) -\tilde{x}(l)$
due to rotation
\begin{eqnarray}
\delta T(\bfl) &=& 0 \,,\\
\delta B(\bfl)      &=& \intlp{}
\Big[ \tilde E(\bfl') \cos 2\varphi_{\bfl'\bfl}
     -\tilde B(\bfl') \sin 2\varphi_{\bfl'\bfl} \Big]W(\bfL)
,\nonumber\\
\delta E(\bfl)      &=& -\intlp{}
\Big[ \tilde B(\bfl') \cos 2\varphi_{\bfl'\bfl}
     +\tilde E(\bfl') \sin 2\varphi_{\bfl'\bfl}\Big]W(\bfL),\nonumber
\end{eqnarray}
where $\bfL=\bfl - \bflp$, and
$\varphi_{\bfl \bflp}= \varphi_\bfl - \varphi_\bflp$ and  $W(\bfL) = 2\alpha(\bfL).$
Thus, due to rotation, a mode of wavevector $\bfL$ mixes the polarization
modes of wavevectors $\bfl$ and $\bf{l^{\prime}}=\bfl -\bfL$. Taking the ensemble average of the CMB fields for the fixed $\alpha$ field,
one gets
\begin{equation}
\langle x^\star(\bfl) x'(\bflp) \rangle_{\rm CMB} =  \langle \tilde{x}^\star(\bfl)\tilde{x}'(\bflp)\rangle
+ f_{xx'}(\bfl,\bflp) \alpha(\bfL)\,.
\label{BasicDifference}
\end{equation}
The $TE$, and $TB$
correlations are produced indirectly via non-zero primordial TE correlation.
\begin{table}[t]
\caption{Minimum variance filters}
\begin{center}
\begin{tabular}{ll}
\hline \hline
$x x'$                & $f_{xx'}({\bfl_1,\bfl_2})$ \vsp \hline $T
E$      & $2 \tilde C_{l_1}^{T E}\sin 2
                  \varphi_{\bfl_1\bfl_2}
                $\vsp
$T B$      & $2 \tilde C_{l_1}^{T E}\cos 2
                        \varphi_{\bfl_1\bfl_2}
                  $\vsp
$E E$           & $2[\tilde C_{l_1}^{E E}
                  -\tilde C_{l_2}^{E E}
                        ]\sin 2
                        \varphi_{\bfl_1\bfl_2}
                        $ \vsp
$E B$           & $2 [\tilde C_{l_1}^{E E}
                 -\tilde C_{l_2}^{B B}]
                \cos 2
                        \varphi_{\bfl_1\bfl_2}
                $
                \vsp
$B B$           & $[\tilde C_{l_1}^{B B}
                 + \tilde C_{l_2}^{B B}]
                        \sin 2
                        \varphi_{\bfl_1\bfl_2}
                $ \vsp
\end{tabular}
\end{center}
\end{table}

Our goal is to use Eq.~(\ref{BasicDifference}) to construct a suitable 
estimator of the  
Fourier components $\alpha(\bfL)$  of the rotation field 
in terms of the observed fields $T (\bfl), E (\bfl), B(\bfl)$ and a
theoretical computation of the power spectra involving un-rotated fields. 
Following ~\citet{2002ApJ...574..566H}, we can define an unbiased quadratic
estimator $\hat{\alpha}_{xx'}(\bfL)$ for $\alpha(\bfL)$  
for each combination of the CMB modes  $x(\bfl_1),x'(\bfl_2)$
by weighting quadratic combinations of different polarization 
modes by $F_{xx'}(\bfl_1,\bfl_2)$ appropriately:
\begin{eqnarray}
\hat \alpha_{xx'}({\bfL})& =&  A_{xx'}(L) \intl{1}
\big[x(\bfl_1) x'(\bfl_2) \nonumber \\ && -\langle \tilde{x}(\bfl_1) \tilde{x}^\prime(\bfl_2)\rangle
\big]
F_{xx'}(\bfl_1,\bfl_2)\,,
\label{eqn:estimator}
\end{eqnarray}
where $\bfL=\bfl_2  -\bfl_1$, and the normalization
\begin{eqnarray}
A_{xx'}(L) = \Bigg[ \intl{1} f_{xx'}(\bfl_1,\bfl_2)
F_{xx'}(\bfl_1,\bfl_2) \Bigg]^{-1} \,,
\label{eq:noise}
\end{eqnarray}
is chosen to make the estimator unbiased, i.e. $\langle \hat\alpha(\bfL)\rangle=\alpha(\bfL)$. 
The fields $x(l)$ can be obtained from the map of an experiment, while the CMB power spectrum of un-rotated but 
 lensed fields can be computed from publicly available
Boltzmann codes like CMBfast and CAMB. 

The weights $F_{xx'}$ can be optimized by minimizing the variance $\langle \tilde \alpha_{xx^\prime}(\bfL) \tilde \alpha^\star_{xx^{\prime}}(\bfL^{\prime})\rangle$. For $x x'=T B$ and $EB$
\begin{equation}
F_{xx'}(\bfl_1,\bfl_2) = \frac{f_{xx'}(\bfl_1,\bfl_2)}{C_{l_1}^{xx} C_{l_2}^{x'x'}}\,,
\end{equation}
         where $C_{l_2}^{xx}$ and $C_{l_2}^{x'x'}$ are the observed power spectra
         including the effects of both the signal and the noise,
\begin{eqnarray}
C^{xx'}_\ell=\tilde C^{xx'}_\ell + C^{xx',n}_\ell\,,
\end{eqnarray}
where $ C^{xx',n}_\ell$ is the noise power spectrum. We assume the detector noise to
be known apriori, be isotropic and Gaussian distributed. We include effects
of beam-smearing by a symmetric Gaussian beam. Then, the noise power spectrum 
is 
\begin{equation}
C_l^{xx,n} = \Delta^2_x e^{l^2\Theta^2_{fwhm}/8 \ln 2 },
\end{equation} 
where $\Delta_x$ is
the instrument noise for temperature (x=T) or polarization (x=P); and $\Theta_{fwhm}$ is the full-width half-maximum (FWHM) resolution of the Gaussian beam. We will assume a fully polarized detector, for which $\sqrt{2}\Delta_T=\Delta_P$.

The variance of the estimator is
\begin{equation}
\langle \tilde \alpha_{xx^\prime}(\bfL) \tilde \alpha^\star_{xx^{\prime}}(\bfL^{\prime})\rangle=(2\pi)^2 \delta(\bfL - \bfLp) \{ C^{\alpha \alpha}_L +N_{xx^\prime}(L)\}\,,
\label{eqn:variance}
\end{equation}
where for the minimum variance estimator, the Gaussian noise $N_{xx^\prime}(L)=A_{xx^\prime}(L)$, and gives the dominant contribution to the variance.

The Gaussian noise
$N(L)$ is dependent only on the instrumental noise power spectrum
$C_l^{x x'n}$ and the power spectrum of the un-rotated polarization field
$\tilde{C}_l^{xx'}.$ Hence the estimator noise depends on the cosmological parameters through the 
power spectrum of the polarization fields. We choose a standard 
 fiducial model with a flat $\Lambda CDM$ cosmology, with no rotation
(i.e. $\alpha =0$), with parameters described by the best fit to WMAP5~\cite{wmap5_cosmology}, given by
$\Omega_b=0.045, \Omega_c=0.23, H_0=70.5, n_s=0.96, n_t=0.0,$ and $\tau=0.08$.

Since, in reality the polarization field will be gravitationally lensed by
the inhomogeneities in the matter distribution in the fiducial model, it
is appropriate to use the power spectrum of lensed anisotropies as the
un-rotated field in calculating $N(L)$. An angular remapping of photon positions due to gravitational 
lensing may  result in an apparent frequency independent rotation 
of the plane of polarization, potentially biasing the estimator. 
Here, we show that the bias is negligible. 
Taking lensing into account the average of the estimator is
\begin{eqnarray}
\langle \hat \alpha_{xx'}({\bfL}) \rangle_{CMB}= \hspace{4.6cm}\nonumber \\ \alpha({\bf L}) + A_{xx'}(L) \intl{1} f^{lens}_{xx'}F_{xx'}(\bfl_1,\bfl_2) \len(\bfL)\,,
\end{eqnarray}
where $\len$ is the line of sight projection of the gravitational potential $\Psi({\bf x})$. The first term on the right hand side is the desired 
rotation field, and the second term represents the bias from 
lensing. The form of lensing filters $f^{lens}_{xx'}$ can 
be found in Table I of~\cite{2002ApJ...574..566H}. The lensing filter $f^{lens}_{xx'}$ are nearly orthogonal to the rotation window 
$F_{xx'}$. Hence the integrand of 
the lensing bias oscillates around zero, and even for $\len\sim 1$, the bias is 
negligibly small in comparison to the square root of 
Gaussian noise $\sqrt{N(L)}$. Since 
Gaussian noise sets the minimum detectable rotation $\alpha(L)$, 
we can neglect the bias for all practical purposes.

\subsection{Rotation from non-standard interactions}
\label{sec:PNGB}
As discussed in introduction, apart from instrumental systematics, which we will discuss in section~\ref{sec:syst}, a way of generating frequency independent rotation is by pseudoscalar fields coupled to photons.

There are no pseudoscalars in the standard model of particle physics 
that couple to radiation. However, they are common 
in particle physics beyond the standard model 
~\cite{1977PhRvL..38.1440P,1978PhRvL..40..279W,1978PhRvL..40..223W,2008LNP...741..199B}. In cosmology they have
been invoked for dark matter or dark energy models, and also as solution to the fine-tuning problem of dark 
energy
~\cite{Frieman:1995pm,Freese:1990rb,Wetterich:1987fm,
Kaloper:2005aj,Dutta:2006cf,Abrahamse:2007te}.

A pseudoscalar field $\phi$ can couple to the electromagnetic
fields by a Chern-Simons interaction term~\cite{prs08,Carroll:1998,2008arXiv0810.0403L} 
\begin{equation}
{\cal L}=-\frac{1}{4}F_{\mu \nu}F^{\mu \nu}+\frac{1}{2}\partial_\mu \phi \partial^\mu \phi + \frac{\phi}{2M}F_{\mu \nu}{\tilde F}^{\mu \nu}\,,
\end{equation}
where $F_{\mu \nu}$ is the electromagnetic field strength tensor, $\tilde F^{\mu \nu}$ is its dual; and the pseudoscalar coupling with electromagnetic field is supressed by mass scale $M$.

The interaction term  $\frac{\phi}{2M}F_{\mu \nu}{\tilde F}^{\mu \nu}$ is invariant under $U(1)$ gauge transformations, 
Lorentz symmetry and parity, and is suppressed by a mass scale M. The fact 
that no such effect has been detected in the laboratories puts a lower bound
on M. It has been shown that such a term can rotate polarization vector of linearly polarized light by an angle of rotation 
$\alpha = \frac{1}{M}\int d\tau \dot{\phi}$ 
during propagation for a conformal time $\tau$~\cite{Carroll98}, which is largest when the 
source of polarization is farthest, which happens for the CMB 
~\cite{1999PhRvL..83.1506L}. 

The angle of rotation along a line 
of sight depends on the change of $\phi$ along that line of sight. We can write the field $\phi$ in terms of a homogeneous piece $\phi_0(\tau)$ and
a position dependent perturbation $\delta \phi({\bn},\tau)$. Then, the rotation angle of the CMB polarization  
\begin{eqnarray}
\alpha({\bn})&=&\frac{1}{M}\Delta \phi({\bn})=\frac{1}{M}\{\phi_0(\tau_0) -\phi_0(\tau_{dec}) -\delta \phi({\bn},\tau_{dec})\}\,,\nonumber\\
\end{eqnarray}
where $\tau_0$ and $\tau_{dec}$ are the conformal times today and at the surface of last scattering respectively, and the perturbation $\delta \phi$ today, at the detector can be taken to be zero without any loss of generality. The shift symmetry of the Lagrangian implies that the field $\phi$ is
classically massless. In our toy example we will assume that this was the case during inflation, and the quantum fluctuations during inflation were frozen in the field and will result in spatial fluctuations today. In this case we can write the power spectrum of fluctuations as nearly nearly scale invariant, 
\begin{eqnarray}
\langle \phi({\bf k})\phi^\star({\bf k'})\rangle &=&(2\pi)^3 P(k) \delta({\bf k} -{\bf k'}) \nonumber\\
&=&(2\pi)^3 c_{\phi} k^{n_s-4}\delta({\bf k} -{\bf k'})
\end{eqnarray} with $n_s=0.96$~\cite{wmap5_cosmology}, and $c_{\phi}=H^2/{2},$ where $H$ is the Hubble parameter during the inflation. Computing the transfer function we  find the power spectrum of the rotation angles to be
\begin{eqnarray}
C^{\alpha \alpha}_{\ell}=\frac{2}{M^2 \pi}\int k^2 dk P_{\phi}(k)j^2_{\ell}(kr)\Delta^2_{\phi}(k,\tau_{dec})\,,
\label{eq:pspowerspectrum}
\end{eqnarray}
where $r=\tau_0-\tau_{dec}$ is the distance to the surface of last scattering, and $\Delta_{\phi}(k,\tau_{dec})=3j_1(k\tau_{rec})/k\tau_{dec}$ is the transfer function which is unity for scales larger than horizon size and decay with an oscillating envelop for modes inside the horizon.

\begin{figure*}[t]
\includegraphics[width=108mm,clip]{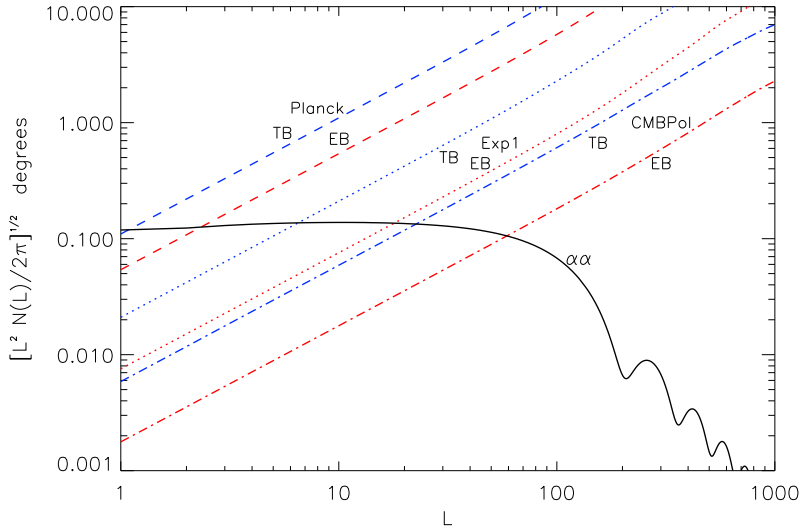}
\caption{
Estimator variance $N(L)$ for the
$EB$ (red curves) and $TB$ (blue curves) estimator as a function of multipole $L$. We show the noise for three experimental setups, CMBPol (dot-dashed), Exp1 (dot) and PLANCK (dashed). The solid black curve shows the rotation angle power spectrum for the model in which the pseudoscalar coupling to the electromagnetic 
field is suppressed by a mass scale $M$~\cite{prs08}, and the perturbations in $\phi$ are seeded during the inflationary phase. We choose the fiducial value of the amplitude for power spectrum $c_\phi/M^2=10^{-4}$. This correspond to energy scale $M\sim 10^{15}\, H_{14}$ GeV, where $H_{14}$ is the Hubble parameter during inflation in units of $10^{14}$ GeV. Note that with CMBPol like experiment, one is sensitive to energy scale as large as $M \sim 10^{17} H_{14}$ GeV, i.e. three orders of magnitude of improvement over the current best constraints. }
\label{fig1}
\end{figure*}

We have discussed the scenario of physical interactions leading to rotation
of the polarization field. There are two important physical parameters 
which determine the magnitude of the effect in this toy model: 
(a) the energy scale (or equivalently the Hubble rate) during 
inflation, which sets the amplitude of the fluctuation power spectrum of
mass-less fields, and (b) The mass scale M by which the Chern-Simons 
interaction term is suppressed. The mass scale M is certainly much
larger than the current energy scales of particle physics experiments 
$\sim 10^2$ GeV, and could be around or higher than the GUT scale 
$\sim 10^{16}$ GeV. In fact, the current best constraints on mass scale $M$ come from the upper limit on B-modes, $M> 2.4\times 10^{14} H_{14}$ GeV~with $H_{14}=H / 10^{14}$~\cite{prs08}. The exact value of the Hubble  scale during inflation is also unknown, with current
estimates suggesting it to be around $\sim 10^{14}-10^{15}$ GeV. For reference the tensor to scalar ratio of perturbations produced in standard models of inflation is given by $r=0.14 \ H^2_{14}$. This for current WMAP5 constraints~\cite{wmap5_cosmology} of $r<0.2$ translates to $H_{14}<1.2$. Since the potential of detection is linked to an assumed 
energy scale of inflation, and M,  it is useful to write rotation angle in terms of $H/M$. The {\it rms} value of rotation angle for $L\lessapprox 100$ can written as $\sqrt{\frac{L^2C^{\alpha\alpha}(L)}{2\pi}}\sim \frac{10H}{M}$ deg. Our approach is to find the magnitude
of the rotation which would be detectable.

\begin{figure}
\begin{center}
\includegraphics[width=88mm,angle=0]{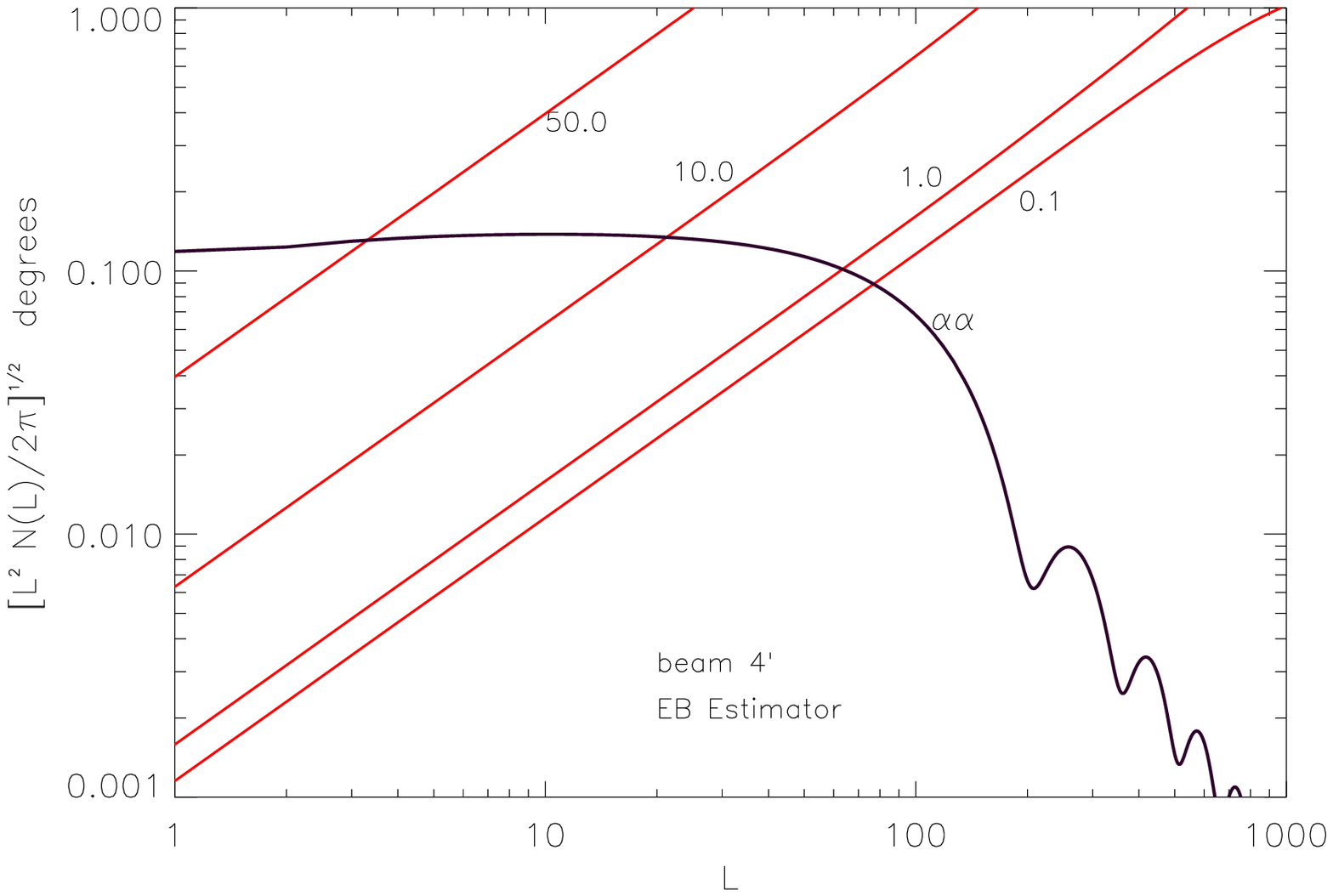}
\includegraphics[width=88mm,angle=0]{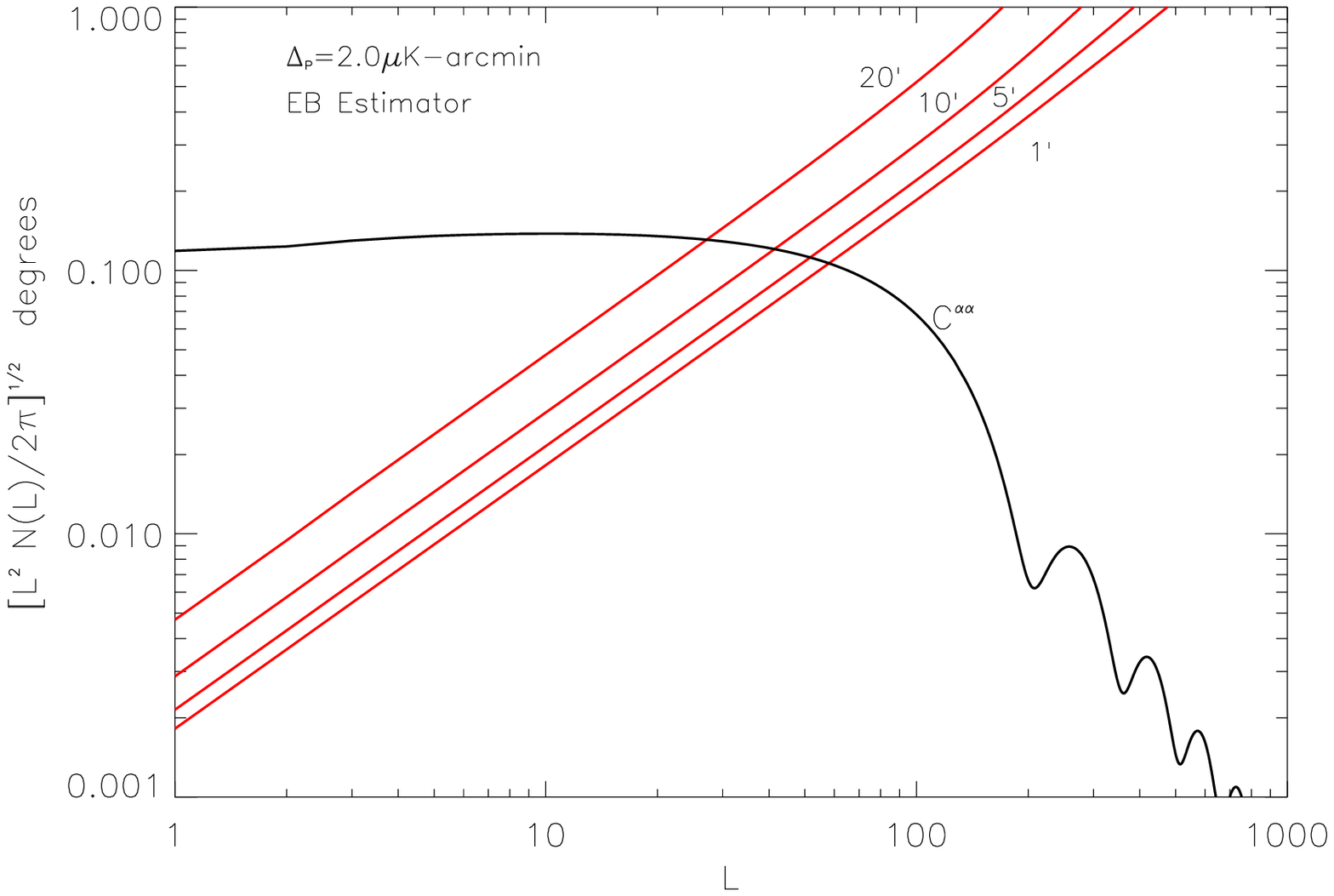}
\caption{Dependence of the variance $N(L)$ of the EB estimator on instrumental 
characteristics as a function of multipole $L$: 
{\it Upper panel}: The diagonal lines represent the variance $N(L)$ for fixed 
FWHM of $4'$
but varying detector moise $\Delta_p$. 
{\it Lower panel}: The diagonal lines represent the variance $N(L)$ for fixed 
detector noise of $\Delta_p=2\mu K$-arcmin but varying the beam size.
In both the panels the oscillatory curve represents the power spectrum of rotational
field for the fiducial amplitude of $c_\phi/M^2=10^{-4}$.}
\label{fig_noise}
\end{center}
\end{figure}

\begin{figure}
\begin{center}
\includegraphics[width=88mm,angle=0]{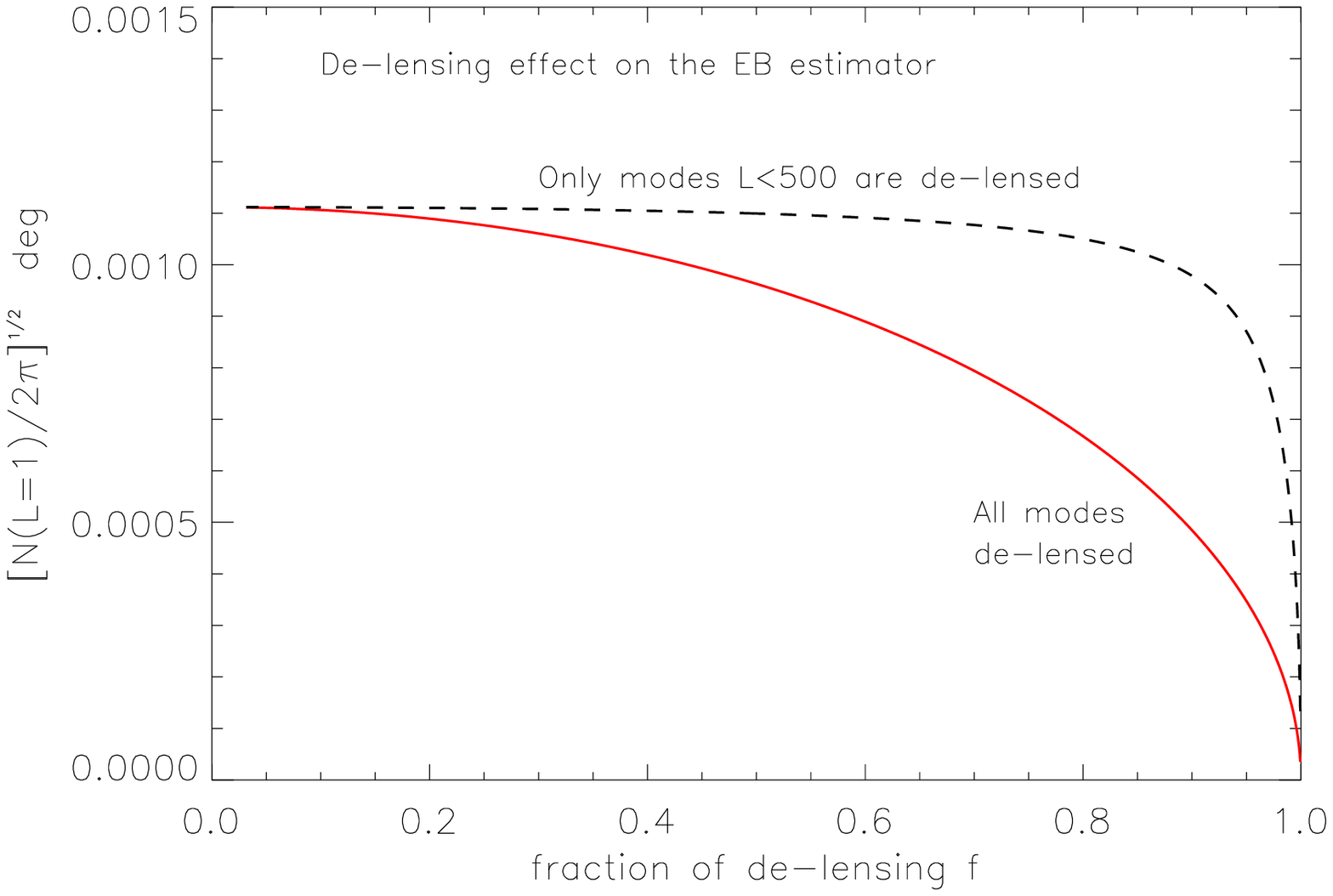}
\caption{Gaussian noise (at L=2) for the EB estimator as a function of fraction of De-lensing of the CMB.  Since the lensing effects are only important for experiments with $\Delta_p < 5 \mu$K-arcmin, we assume an idealized experiment (i.e. $\Delta=0, \Theta_{fwhm}=0$) to see the effect of de-lensing. Lower (solid red) curve assumes that all the observed B-modes ($l_{max}=3000$) are being de-lensed by equal amount. Upper (dash black) curve shows the effect of de-lensing when only the modes with $\ell<500$ are being de-lensed, while no de-lensing for $\ell>500$.}
\label{fig:delens}
\end{center}
\end{figure}

\section{Results}


\subsection{Estimator}
\label{Estimator}
We discuss the prospects of using our unbiased estimator 
$\hat{\alpha}(\bn)$ for detecting Fourier modes 
of the spatially varying rotation $\alpha(\bn)$. We study the estimator variance and its dependence on instrumental characteristics 
focusing on planned missions.

We want to find out the magnitude of rotation that would be detectable. For this, the rotation signal must be 
larger than the noise in the estimator $N_{xx'}(L)$. 
In Fig.~\ref{fig1}, we show the variance for the $EB$ (red curves) and $TB$ (blue curves) estimators 
\footnote{For the three experiments considered in Fig.~\ref{fig1}, 
the other estimators have much smaller sensitivity to the 
rotation angle.} 
as a function of multipole $L$ for the three experimental setups, 
(1) PLANCK satellite with noise level $\Delta_P=56\mu K$-arcmin 
and FWHM of 7',
(2) experiment with noise level $\Delta_P=9.6 \mu K$-arcmin and 
FWHM of $8'$, typical of upcoming ground and balloon-based CMB 
experiments (hereafter called Exp1), and
(3) a CMBPol-like instrument with noise level 
$\Delta_P=\sqrt{2}\mu K$-arcmin and FWHM of $4'$, typical of 
future space-based CMB experiments. The black solid line shows the power spectrum of the cosmological rotation signal $C^{\alpha\alpha}_l$ from the Chern-Simon coupling discussed in section~\ref{sec:PNGB}. We show this signal for
$c_\phi/M^2=10^{-4}$, which corresponds to $M= 1.5\times 10^{15}\, H_{14}$ GeV.

For the three experiments in consideration, the $EB$ estimator is found to be the most 
sensitive, and the  
variance is roughly constant up to  $L\sim1000$. Although, we do not show  $L=0$ 
in the plot, our estimator can also be used to estimate the detectability of uniform rotation. For uniform rotation, $\bfL=\bfl_1-\bfl_2=0$ in Eq. (\ref{eq:noise}), hence there is no mode mixing between different wavevectors. The variance $N(L)$ for monopole and dipole 
are comparable; so rotation $\alpha(L)$ can be constrained to 
similar levels for these modes.

 Currently only the constant angle of rotation (i.e. $L=0$ of our estimator) has been constrained, $\alpha < 2^o$ ~\cite{quad_parity,wmap5_cosmology}; and at present there are no constraints on the spatial variation of the rotation angle  $\alpha({\bf n})$. From our Fig.~\ref{fig1} upcoming experiment like PLANCK will be able to detect the rotation angle power spectrum $C^{\alpha \alpha}_\ell$ as small as $0.01$ deg$^2$, while futuristic experiment like CMBPol will be able to detect rotation angle power spectrum as small as $2.5 \times 10^{-5}$ deg$^2$. These numbers translate to minimum detectible $H/M=4\times 10^{-3}$ for PLANCK, and $H/M=2\times 10^{-4}$ for CMBPol. 

\begin{figure*}[t]
\begin{center}
\includegraphics[width=88mm,angle=0]{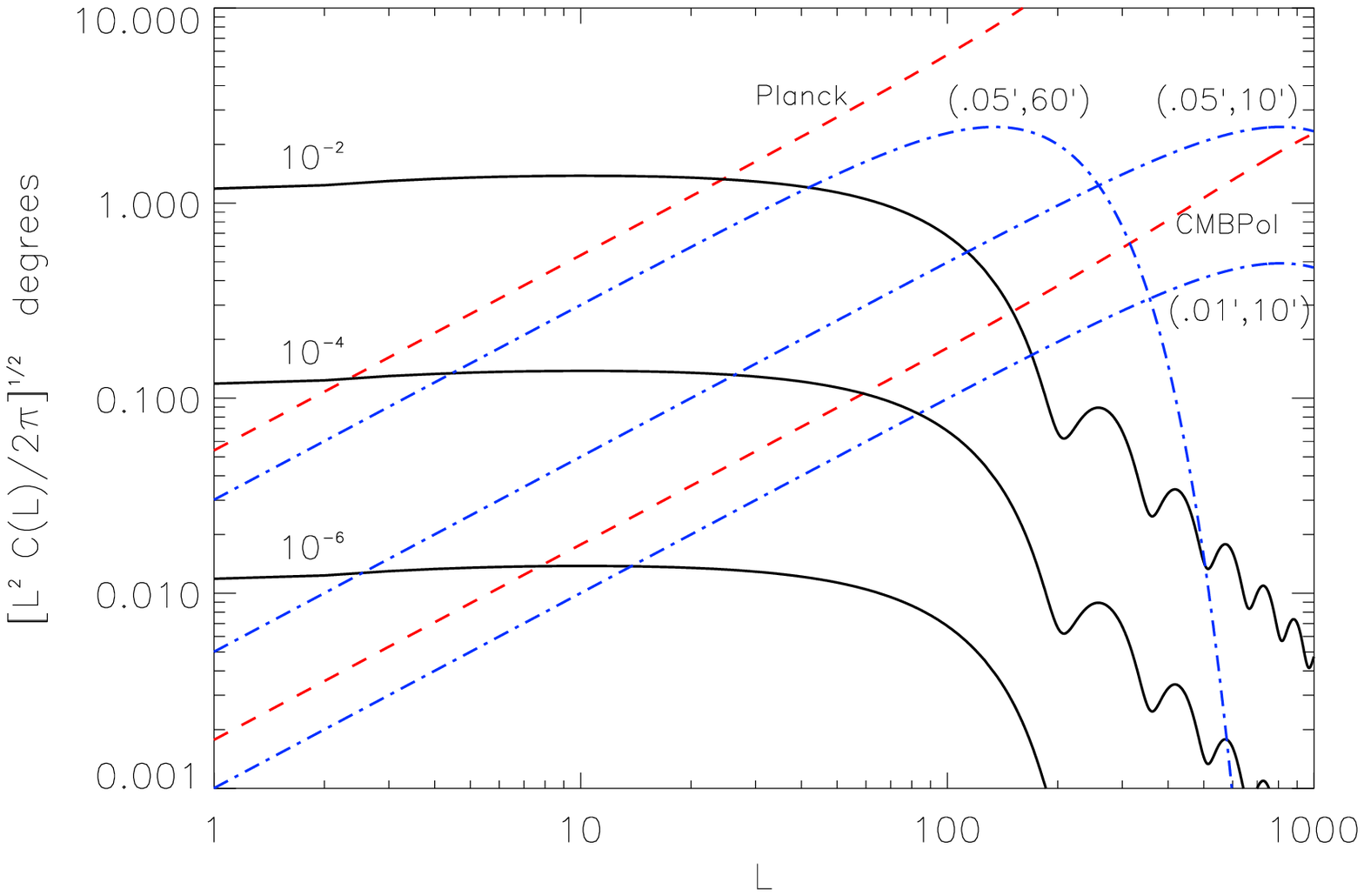}
\includegraphics[width=88mm,angle=0]{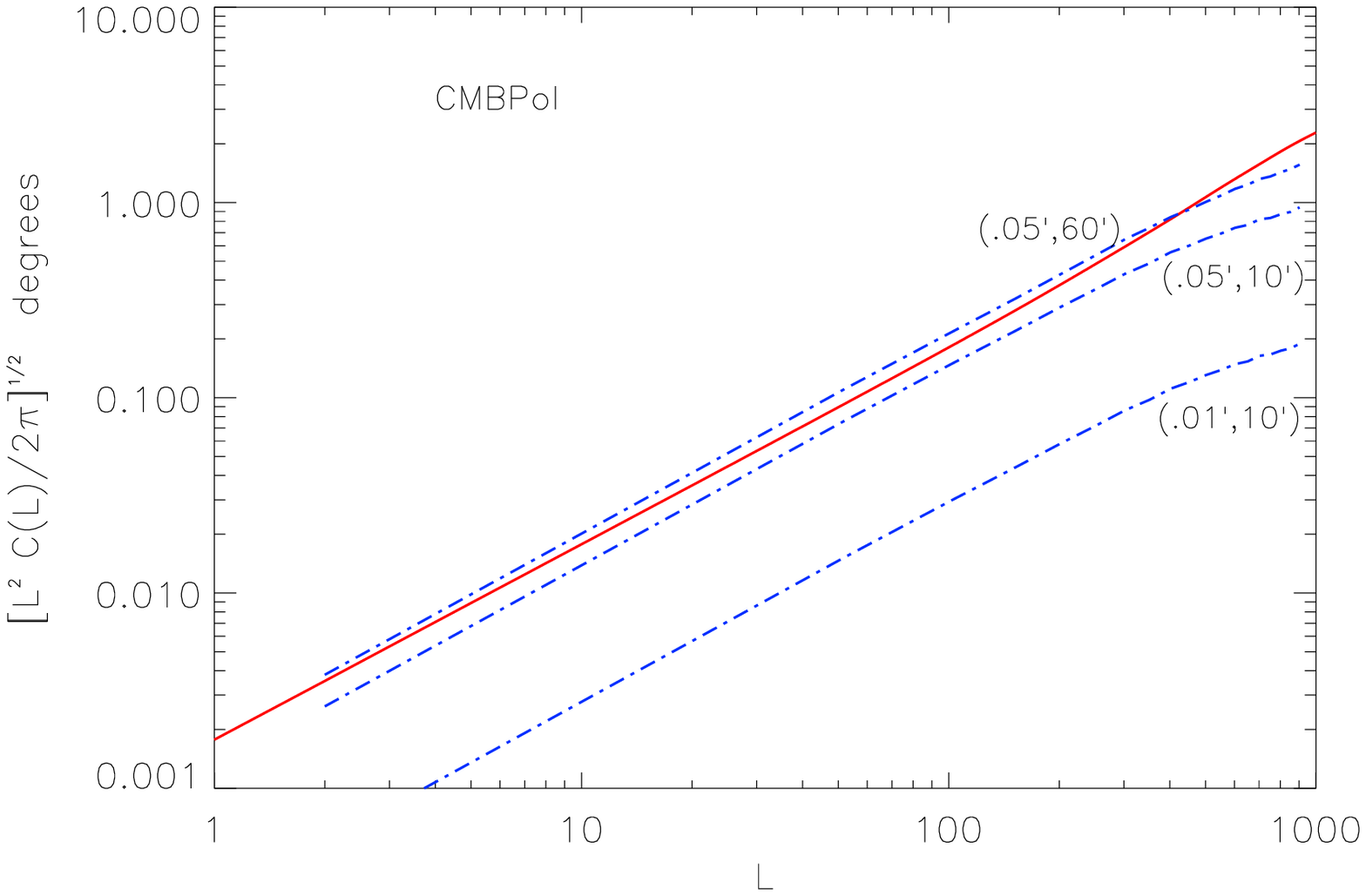}
\caption{{\it Left Panel}: Dot dashed (blue) curves show 
the systematics rotation power spectrum 
$C^{\omega \omega}(L)$ for three combinations of {\it rms} amplitude 
$A_\omega$ (in arcmin) and FWHM of coherence length, $\sqrt{8\ln(2)}\sigma_\omega$ 
(in arcmin). The two dashed red curves show the 
Estimator variance 
$N_{EB,EB}(L)$ for the PLANCK (upper) and CMBPol (lower) 
experiment. The black oscillatory curve shows the power spectrum of rotational
field for the fiducial amplitude of $c_\phi/M^2=10^{-2},10^{-4},$ and $10^{-6}$. {\it Right Panel}: For 
CMBPol experiment, solid red curve shows the estimator variance
$N_{EB,EB}(L)$ and the dot dashed (blue) curves show the systematics 
noise $N^{\omega \omega}_{EB,EB}(L)$ for three combinations of {\it rms} amplitude 
$A_\omega$ (in arcmin) and FWHM of coherence length, $\sqrt{8\ln(2)}\sigma_\omega$ 
(in arcmin) as the left panel.}
\label{fig:syst}
\end{center}
\end{figure*}

In Fig.~\ref{fig_noise}, we study the dependence of the 
estimator variance with the detector noise $\Delta$, and 
the beam size $\Theta$. As in Fig.~\ref{fig1}, here also we show the cosmological signal $C^{\alpha \alpha}(L)$ for reference, with the same fiducial parameter $c_\phi/M^2=10^{-4}$. 
The upper panel shows $N_{EB,EB}(L)$ as a function of $L$ for various choices of instrumental noise $\Delta_p$ but for a fixed beam size of $4'$.
The lower panel shows estimator noise $N_{EB,EB}(L)$ as a function of $L$ for $\Delta_p=2\mu K$-arcmin and for various choices of  beam size, starting from $20'$ to $1'$.  For FWHM of the beam size $\Theta_{fwhm}\sim 5'-50'$, and instrumental noise $\Delta_p \sim 5-50 \mu K$-arcmin, the scaling of variance $N_{EB,EB}(L)$ can be fitted by a simple power law $N_{EB,EB}(L)=3.3\times 10^{-7} \Delta^2_p \Theta^{1.3}_{fwhm}$ deg$^2$. However, we cannot minimize the estimator noise $N(L)$ to 
arbitrarily low levels by reducing the detector noise; the estimator variance plateaus out at a level of $\Delta_p < 5 \mu K$-arcmin (larger than 
detector noise levels in CMBPol) to 
$\approx 10^{-6}$deg$^2$. Physically, this is due to lensing effects. 
 
As discussed in Sec.~\ref{sec:formalism_estimator}, the estimator variance $N_{xx',xx'}(L)$ shown in Fig.~\ref{fig1} 
includes contribution  due to lensing of E modes to B-modes.
For experiments like Exp1 and PLANCK with $\Delta_p > 5 \mu$K-arcmin,
lensing effects are negligible compared to the detector noise.
Therefore, there is no difference in the estimator variance if the lensed 
polarization fields used as the un-rotated fields in Fig.~\ref{fig1}, are 
replaced with un-lensed polarization fields. On the other hand,
for experiments with $\Delta_p < 5 \mu$K-arcmin (like CMBPol), 
lensing of the CMB power spectrum dominates the estimator variance,
and eventually limits the sensitivity of the an idealized
instrument to $\sim 10^{-6}$deg$^2$.
Further, we calculated the leading order lensing contribution to
noise, $N^{lens.}(L)$ which is proportional to lensing power spectrum
$C^{len}(L)$, and is related to the connected part of the trispectrum.
We find that this noise $N^{lens.}(L)$  is smaller than the
estimator noise $N_{xx',xx'}(L)$ shown in Fig.~\ref{fig1} for all cases.

{\it Can Estimator noise be further reduced?} Lensing B-modes can be measured and hence, in principle, can be separated (de-lensing) from the pure rotation $\alpha(\bn)$
considered above. In the absence of lensing, the sensitivity of the
idealized instrument would be limited by the cosmic variance of primordial
B-modes.
While de-lensing can improve the sensitivity of the idealized instrument, to the level of the noise inherent
in the de-lensing process, it is likely to be challenging~\cite{CMBPol:lensing}. In Fig.~\ref{fig:delens} we show how much the Gaussian noise for the $EB$ estimator $N(L)_{EB,EB}$ reduces as a function of amount of de-lensing. The de-lensed B-modes
\begin{equation}
B^{de-lens}_\ell = f B^{lens}_\ell\,
\end{equation}
are used in the estimator which reduces the variance, depending on the fraction of de-lensing $f$. 
The lower curve (solid red) shows the most optimistic scenario where all the CMB modes are being de-lensed (up to $L=3000$). The upper curve shows, although still challenging, a more realistic case where only the modes $L<500$ are being de-lensed.

\subsection{Detectability of cosmological rotation and instrumental systematic effects}
\label{sec:syst}
As indicated before, a cosmological rotation field $\alpha({\bf n})$ can be confused with the instrumental rotation systematics.
A calibration error in the angular position of the instrument is
degenerate with
a spatially constant angle of rotation (L=0 of $\alpha(\bfL)$),
while errors in the rotation calibration of individual detectors in the instrument, leading
to a relative mis-alignment of axes of the individual detectors by
angles $\omega_i$ are degenerate with spatially varying
$\alpha(\bn)$.  For an instrument with a large number of detectors, we can
treat the angles of rotation of the i$^{th}$ detector $\omega_i$ as a
smooth field as a function of the detector position. The relative offsets in the polarimeters in the detector could result in a systematics signal $\omega({\bf n})$ in the map if the weighting of each polarimeter changes from pixel to pixels in the map. This depends on the scan strategy. 
For illustration purposes we can model the statistical properties of systematics signal~\cite{HHZ,SYZ09}
as a statistically isotropic Gaussian field with a power spectrum given by
\begin{equation}
C_l^{\omega\omega} = \frac{A^2_\omega \exp(-l(l+1)\sigma_{\omega}^2)}{\int {d^2 l \over (2\pi)^2}
\exp(-l(l+1)\sigma_{\omega}^2)},
\label{eq:coh}
\end{equation}
where $A_\omega$ characterizes the {\it rms} value of this field $\omega$, and $\sigma_\omega$ is a coherence length.

To see the effect of rotation systematic field, we can change $\alpha(\bn)$ in Eq. (\ref{RotationTransformationOfStokes}) to $\alpha(\bn)+\omega(\bn),$ and re-derive our estimator. The systematics field biases our estimator 
$\hat{\alpha}(\bfL)$  by an amount $\omega(\bfL)$ and increases the 
variance of the $xx'$ estimator by an amount 
$C^{\omega\omega}(L) + N^{\omega\omega}_{xx',xx'}(L)$ (see appendix).  For our model of systematics field and assuming that $\alpha$ and $\omega$ fields are uncorrelated, the power spectrum of bias is given by Eq. (\ref{eq:coh})  i.e. $C^{\omega\omega}(L)$. For the expression for the systematic noise $N^{\omega\omega}_{EB,EB}(L)$ please refer to the appendix. In order 
to use the estimator for detection of a rotation field with  power spectrum
$C^{\alpha\alpha}(L),$ $C^{\alpha\alpha}(L) >> 
C^{\omega\omega}(L)$. 

In order for the cosmological rotation field $\alpha(\bn)$ to be determined to the noise levels in Fig.~\ref{fig1}, both the systematic bias power spectrum $C^{\omega \omega}(L)$ and systematic noise $N^{\omega \omega}_{xx',xx'}(L)$ should  be smaller than the estimator noise $N_{xx',xx'}(L)$.
In Fig.~\ref{fig:syst}, we study the dependence of the  systematic field 
power spectrum  $C^{\omega\omega}(L)$ and the systematics noise term 
$N^{\omega\omega}_{EB,EB}(L)$.

In the left panel of Fig.~\ref{fig:syst} the dot dashed (blue) curves show the systematics rotation signal for various choices of {\it rms} amplitude $A_\omega$ and FWHM of coherence length $\sqrt{8\ln(2)}\sigma$. The systematics signal does not depend on the instrumental noise $\Delta$, and beam, $\Theta_{fwhm}$. Solid (black) curves show the fiducial cosmological rotation power spectrum $C^{\alpha \alpha}(L)$ for $c_{\phi}/M^2=10^{-2},10^{-4}, \& 10^{-6}$, and the two dashed (red) curves show the Gaussian noise for the $EB$ estimator $N_{EB,EB}(L)$ for PLANCK (upper) and CMBPol (lower) experiments. 

In right panel of Fig.~\ref{fig:syst} the dot dashed (blue) curves show the systematics noise for EB estimator $N^{\omega \omega}_{EB,EB}(L)$ for the CMBPol like experiment and for various choices of {\it rms} amplitude $A_\omega$ and coherence length $\sigma$. The solid (red) line shows the Gaussian noise for the $EB$ estimator $N_{EB,EB}(L)$ for CMBPol experiment.

\begin{figure*}[t]
\begin{center}
\includegraphics[width=88mm,angle=0]{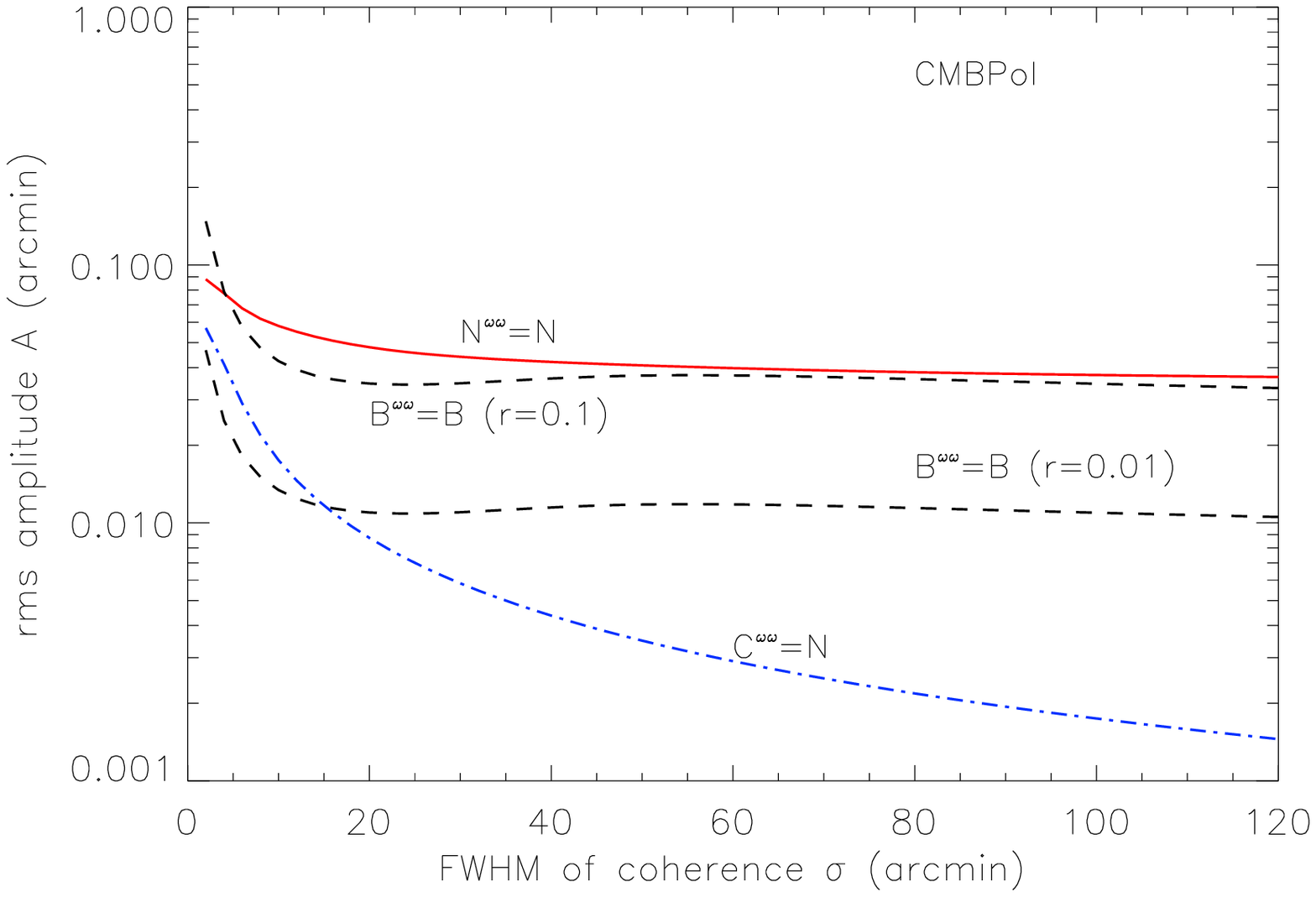}
\includegraphics[width=88mm,angle=0]{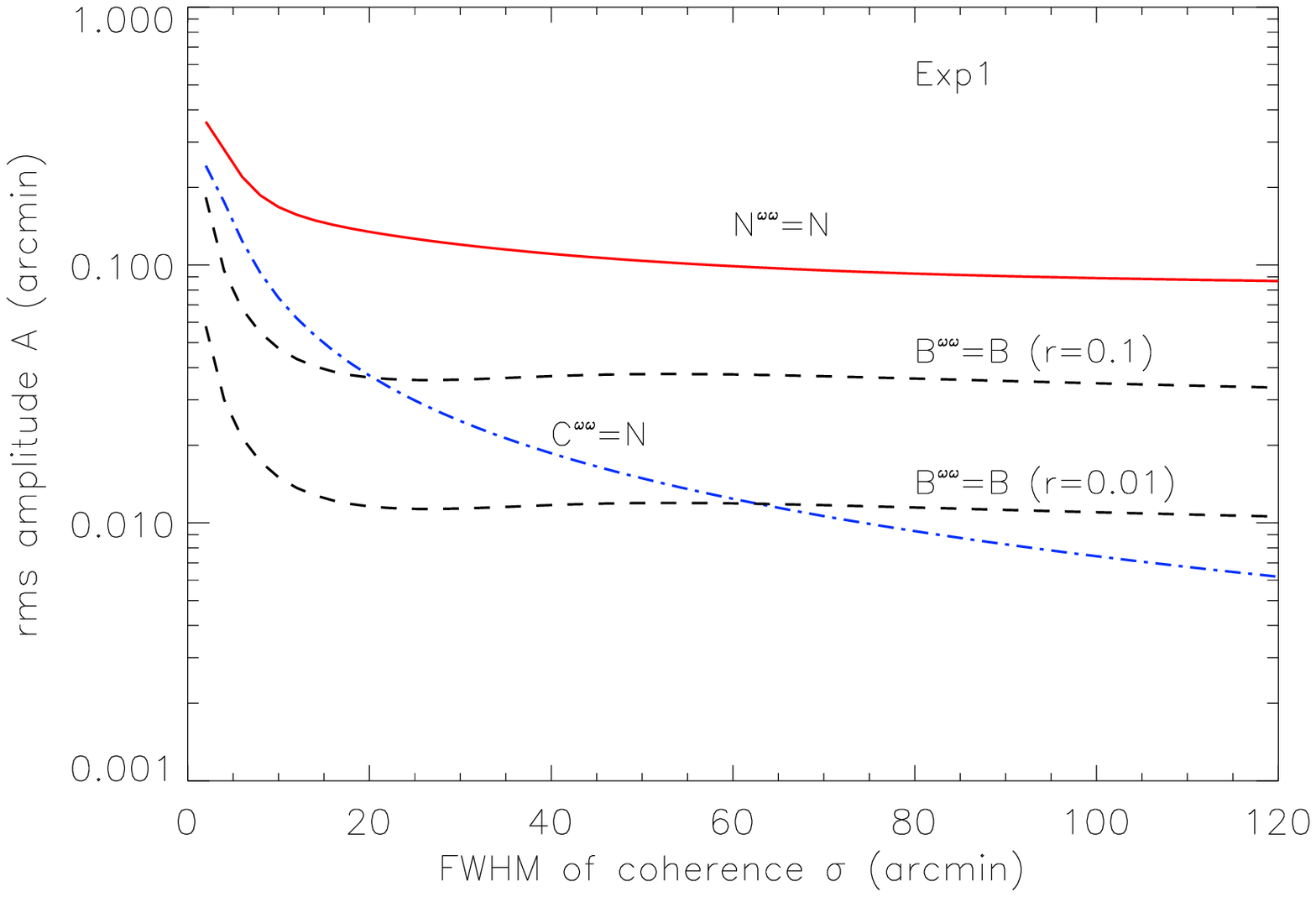}
\caption{The requirement on control of 
the systematic fields $\omega$ for CMBPol ({\it Left Panel}), and Exp1 
({\it Right Panel}) for the detection of primordial B-modes, and the level to which the CMB can be de-rotated.. The dashed black line shows the systematic parameters $\{A,\sigma\}$ which generate spurious B-modes of the same magnitude at ($L=90$) as the primordial B-modes (at L=90, where the primordial B-modes peak); for values of $r=0.1$ (upper back) and $r=0.01$ (lower black). The solid red curve represents the systematic fields for which the variance contribution due to the systematic field $N^{\omega \omega}_{EB,EB}(L=2)$ becomes equal to the estimator variance $N_{EB,EB}(L=2)$. The blue line represents the systematic fields for which the power spectrum of the $\omega$ field at $L=2$ is equal to the estimator variance $N_{EB,EB}(L=2)$. }
\label{fig:systcomp}
\end{center}
\end{figure*}

Note that like the Gaussian noise $N_{EB,EB}(L)$, the systematic noise $N^{\omega\omega}_{EB,EB}(L)$ is also weakly dependent on $L$. However the systematic power spectrum (depending on the coherence length) may only be similarly flat only
 to about $L\sim 100$ (for coherence $60'$).
Hence, we may characterize their values by their values at a particular 
value of $L < 100$. 
Together, the two panels show that at a coherence length of $10'$, an 
{\it rms} amplitude of $\approx 0.01'$ gives a systematic variance which is about 
ten times smaller the estimator variance in the CMBPol experiment, 
while  the bias power spectrum is only about half the estimator variance. 
Smaller coherence length $\sigma$ and {\it rms} amplitude of the systematic 
fields result in smaller effect of the rotation systematics. 
This implies, that in order to detect rotation, using the CMBPol 
experiment, one would have to control the systematic field to much better 
than  $\{\sigma_\omega, A_\omega\}$ values $\{10', 0.01'\}$.

\subsection{De-rotating CMB to improve the sensitivity for the Primordial B-modes Detection}
\label{sec:Bsyst}
An important design goal of the futuristic
CMB polarization experiments is the detection of primordial B-modes. Rotation generates B-modes (via Eq. (4)) which can be confused with the primordial B-modes. Both the instrumental rotation systematics and any cosmological rotation will generate B-modes. Hence, in order to study primordial B-modes, it is necessary to know the level of these spurious 
B-modes.  

 An important application of our estimator is to measure rotation and then in turn de-rotate the CMB polarization field to remove the spurious B-modes and hence increase sensitivity to the primordial B-modes detection. For this application, it is not important to know what the source of this rotation is.

A specific example is the case when cosmological rotation is known (or assumed) to be small and we are interested in controlling instrumental systematics to detect primordial B-modes. One can in this case layout specifics on what is the minimum B-modes amplitude that will be detectable without worrying about rotation systematics. The amplitude of the primordial B-modes is fixed by amplitude of tensor perturbations 
which depends on the energy scale of inflation. Equivalently one can use the ratio of amplitude of tensor and scalar perturbations $r$ to characterize the B-modes.  

In Fig.~\ref{fig:systcomp} we show the comparison of required control of 
the systematics $\omega$ for CMBPol (left panel), and Exp1 (right panel) for the detection of primordial $B$-modes, and the level to which the CMB can be de-rotated. The dashed black lines 
show the $A_\omega$, and $\sigma_\omega$ of the systematic fields (see Eq.~(\ref{eq:coh})) for which spurious $B$-mode power spectrum $C^{\omega\omega}(L)$ at $L=90$ is equal the primordial $B$-modes power spectrum (at L=90) for values of $r =0.1$ (upper curve), and $r=0.01$ (lover curve). 
The solid red curve shows the $A_\omega$, and $\sigma_\omega$ of the systematic field for which the variance contribution due to the systematic field $N^{\omega\omega}_{EB,EB}(L)$ at $L=2$ becomes equal to the Gaussian noise $N_{EB,EB}(L)$ at $L=2$. 
The dot-dashed blue curve shows the $A_\omega$, and $\sigma_\omega$ of the systematic field for which 
the power spectrum of the $\omega$ field $C^{\omega \omega}(L)$ at $L=2$ is equal to the estimator 
variance $N^{\omega \omega}_{EB,EB}(L)$ at L=2. 

In order to be able to detect $B$-modes of a particular $r$ value,
the parameters  $\{A,\sigma\}$ must be in the region below the corresponding black lines.
Dot-dashed blue curves represent the level of rotation systematics to which our estimator can be used to correct for; i.e. if the systematics, $\{A,\sigma\}$ are above the blue curve then our estimator can reduce it to the level given by blue curve. 
If the instrumental systematics parameter $\{A,\sigma\}$ are above the red curve, our estimator is no longer a minimum variance estimator, and one should include the systematics noise effect $N^{\omega\omega}$ in the variance of our estimator. However if instrumental systematics parameter $\{A,\sigma\}$ are below the red curve, then for a given $r$ if the systematics requirement for B-mode is less stringent that the requirement from rotation, one can use our estimator to control the rotation systematics for B-modes detection. 

If the systematics parameters  $\{A,\sigma\}$ are below the dot-dashed blue curve, then the rotation from the systematics is smaller than the sensitivity of the estimator. If our estimator detects rotation signal in this case then it can be attributed to cosmological signal or to incorrectness of the systematics model.


\section{Summary and Discussion}
There are interesting physical mechanisms that can rotate the plane of
CMB polarization.
We presented explicit
formulae for estimators of
the spatially varying rotation angle $\alpha(\bn)$ that
can be constructed from future datasets in the flat sky limit.
By computing the variance of these estimators, we estimate how large a variation in the
angle
$\alpha$ must be to be detected by a particular experiment. Currently only the constant angle of rotation (i.e. $L=0$ of our estimator) has been constrained, $\alpha < 2^o$ ~\cite{quad_parity,wmap5_cosmology}. At present there are no constraints on the spatial variation of the rotation angle  $\alpha({\bf n})$. From our Fig.~\ref{fig1} upcoming experiment like PLANCK will be able to detect the rotation angle power spectrum $C^{\alpha \alpha}_\ell$ as small as $0.01$ deg$^2$, while futuristic experiment like CMBPol will be able to detect rotation angle power spectrum as small as $2.5 \times 10^{-5}$ deg$^2$. 
These numbers translate to minimum detectible $H/M=4\times 10^{-3}$ for PLANCK, and $H/M=2\times 10^{-4}$ for CMBPol.

Gravitational lensing does not bias the estimator, however it increase the variance of the estimator.
The increase in the variance is sub-dominant for experiments with $\Delta_p > 5 \mu$K-arcmin.
For small instrumental noise $\Delta_p \leq 5 \mu K$-arcmin, the lensing B-modes become important, saturating the variance to $\sim10^{-6}$deg$^2$ even for an ideal experiment.

The physical mechanisms that give rise to the rotation field all probe
interesting cosmological physics; in principle, they can be separated using
their frequency dependence and used to study the magnetic
field or polarization dust maps. A cosmological source for a frequency
independent pure rotation field can be interpreted as exotic feature
signifying a clear departure from standard model physics. Such a departure
could be a violation of the equivalence principle, or violation of
Lorentz invariance~\cite{sme,2007astro.ph..2379K}. In the context of cosmology, an important
example is pseudoscalar fields which have been proposed as
dynamical models of dark energy to solve the cosmological constant problem, dark matter, and also as a solution to the fine tuning problem of dark energy.

We have considered a model in which the perturbations in the scalar field were imprinted during inflation and the scalar field couples to photon via Chern-Simons coupling discussed in Sec.~\ref{sec:PNGB}, which are suppressed by mass scale M. With CMBPol like experiment one can constrain the mass scale $M>10^{17} H_{14}$ GeV, where $H_{14}$ is the Hubble parameter during inflation in the units of $10^{14}$ GeV.   

If the scalar field is assumed to be responsible for the dark energy, the constant rotation would
probe the coupling scales of such a dark energy field, and establish its
dynamic nature. The spatially varying part of this rotation field would
also probe the clustering of such a field. Typically, such fields would
have large sound speeds, so that clustering is only possible at large
scales. Hence, the possibility of detection of the spatially varying field
is best at low multipoles. If one assumes that the field has the kind of
clustering discussed in~\cite{2008arXiv0810.0403L}, then from an
experiment like CMBPol, one can constrain the mass scale of suppression of
the Chern-Simons coupling term, to $M \gtrsim 10^{10}$ GeV. 

However, the rotation field induced by a cosmological source can be
degenerate with the rotation systematics of the instrument, which
 are limited by the rotation calibration of the polarimeters.
Thus, the detection of such cosmological signals is only possible if the
cosmological signal is larger than the level of rotation systematics signal that can be
controlled. Rotation systematics also effect the variance of our estimator. We quantify the level of systematics control
required for detection of the cosmological signal to be only limited by
the estimator noise $N(L)$. If one can show from other experiments that the
sources of such cosmological
signals can be limited to magnitudes $\vert \delta \alpha(\bfL)\vert $ smaller than
these systematic levels, then we cannot detect the cosmological signals.
However, then we can use this fact that the observed rotation field should be
less than this magnitude to calibrate the instrument to control the level
of rotation systematics to precision levels of
$\sim \vert \delta \alpha(\bfL)\vert $. This could enable a better
study of effects like primordial B-modes, lensing, or the frequency dependent signals from
sources like magnetic fields or foreground dust. Thus precise studies of
this rotation field could either probe exciting physical effects, and/or
 enable better control of calibration and instrumental statistics.

\acknowledgements{We thank Shaul Hanany, Carlo Baccigalupi, Nicolas Ponthieu, Julian Borrill, Sam Leach and Britt Reichborn-Kjennerud for useful discussions during the project. We especially thank Shaul for discussions which initiated this project. RB would like to acknowledge support from NSFAST07-08849}



\appendix
\section{Contamination} Instrumental rotation systematics and lensing of the CMB can effect our estimator. In this appendix we will show how the rotation systematics and lensing of CMB appear in our estimator. Let us denote systematics field by $\omega(L)$ and lensing filed by $\len(L)$. We can incorporate the effect of rotation systematics by changing $\alpha(\bn)$ to $\alpha(\bn)+\omega(\bn)$, and effect of lensing by changing $\bn$ to $\bn+d(\bn)$ in Eq. (1), where ${\bf d}=\nabla\len$. In the presence of rotation systematics and lensing, the average of the estimator is given as
\begin{eqnarray}
\langle \hat \alpha_{EB}({\bfL}) \rangle &=&  \alpha({\bf L}) + \nonumber \\
&&\omega({\bf L}) + A_{EB}(L) \intl{1} f^{len}_{EB}F_{EB} \phi(L),\nonumber \\
\end{eqnarray}
where the first term on the right hand side is the desired rotation
field, the second term is the estimator bias from instrumental
systematics, and the third term represents the bias from lensing. 
The variance of the estimator can be written as
\begin{eqnarray}
&&\Big\langle\Big\langle \langle \estEB(\bfL) \cdot \estEB(\bfLp)
\rangle_{\rm{CMB}}} \Big\rangle_{\rm{LSS}}\Big\rangle_{\rm{SYS} \nonumber \\
\quad &=&{\normEB(L)}{\normEB(L')}\times \nonumber \\
&&\quad\intl{1} \intl{1'}
\filtEB(\vecla, \veclb) \filtEB(\vecla', \veclb') \nonumber \\
&&\quad \times \Bigg\{\left< E(\bfl_1)^\obs
B(\bfl_2)^\obs E(\bfl_1')^\obs B(\bfl_2')^\obs\right>
\Bigg\} \nonumber \\
\quad &=& (2 \pi)^2 \delta_\dirac(\bfL + \bfLp)\times \nonumber \\
&&\Bigg[
C^{\alpha\alpha}(L) + C^{\omega\omega}(L) + \noiseEB(L) +\noiseAEB(L) \nonumber \\
&&\quad  + \noiseWEB(L) + \noiseLEB(L)+...\Bigg] \,,
\label{eq:varianceEB}
\end{eqnarray}
where $\bfL=\vecla + \veclb$, $C^{\alpha\alpha}(L)$ is the cosmological rotation
power spectrum, and $C^{\omega\omega}(L)$ is the rotation
systematics power spectrum. The terms $\noiseEB(L)$, $\noiseAEB(L)$,
$\noiseWEB(L)$, and $\noiseLEB(L)$ are the estimator Gaussian noise,
the first order non-Gaussian estimator noise, the first order systematics noise of
instrumental rotation, and the first order lensing induced
non-Gaussian noise, respectively. Like the lensing quadratic estimator, the
Gaussian noise comes from the disconnected part of the four-point
function, while non-Gaussian noise $N^{(\alpha)}_{EB,EB}$,
$N^{(\omega)}_{EB,EB}$, and $N^{(len)}_{EB,EB}$ comes from the
connected part. We note that the Gaussian noise term also includes
rotation systematic effects implicitly since instrumental
systematics bias the measured rotation power spectrum with
$C^{\omega\omega}(L)$ with as we have shown, note that we assume no
cross correlation term $C^{\alpha\omega}(L)$. The ellipses in
Eq.~(\ref{eq:varianceEB}) stands for higher order terms. The first order non-Gaussian noise can be written as
\begin{eqnarray}
\noiseXEB(L)=\hspace{6cm} \nonumber \\
{\normEB^2(L)}\intl{1} \intl{2} \filtEB(\vecla, \veclb)  \filtEB(\vecla', \veclb') \times \nonumber \\
\Bigg[ C_{l_1}^{EE}C_{l_1'}^{EE} \Bigg\{
C_{|{\vecl_1}+{\vecl_2}|}^{XX}W^{X}_{B}(\vl_2,-\vl_1)W^{X}_{B}(\vl_2',-\vl_1')\,\,\,\,\,\,\,\, \nonumber \\
+
C_{|{\vecl_1}+{\vecl_2'}|}^{XX}W^{X}_{B}(\vl_2,-\vl_1')W^{X}_{B}(\vl_2',-\vl_1)\Bigg\} \Bigg]\,, \hspace{.5cm}
\end{eqnarray}
where $X = \{\alpha, \omega, \len\}$, the window functions
$W^{\omega}_{B}(\bfl_1,\bfl_2)=W^{\alpha}_{B}(\bfl_1,\bfl_2) = 2 \cos[2 (\varphi_{l_2} -
\varphi_{l_1})]$, and $W^{\len}_{B}(\bfl_1,\bfl_2) = \sin 2
(\varphi_{\bfl_1} -\varphi_{\bfl_2})(\bfL \cdot \bfl_1)$. We use
this equation to numerically compute the systematic-induced
estimator noise for the rotation systematics. Among these extra
covariance noise, $N^{(\alpha)}_{EB,EB}$, and $N^{(L)}_{EB,EB}$ are
always smaller than the estimator Gaussian noise $\noiseEB(L)$, and
$N^{(\omega)}_{EB,EB}$ can be in some cases comparable to
$\noiseEB(L)$. 
We use Fig.~\ref{fig:systcomp} to illustrate when
$N^{(\omega)}_{EB,EB}$ goes to the same level as $\noiseEB(L)$
under certain experiment configuration.

\end{document}